\documentstyle[epsf]{mn}

\title[Galactic warps in prolate haloes]
{Time evolution of galactic warps in prolate haloes}
\author[M.~Ideta et al.]
{M.~Ideta,$^{1}$\thanks{E-mail: ideta@kusastro.kyoto-u.ac.jp (MI);
    hozumi@sue.shiga-u.ac.jp (SH); tsuchiya@kusastro.kyoto-u.ac.jp (TT);
    takizawa@kusastro.kyoto-u.ac.jp (MT)}
  S.~Hozumi,$^{2\,{\mbox{\LARGE$\star$}}}$
  T.~Tsuchiya$^{1\,{\mbox{\LARGE$\star$}}}$
  and M.~Takizawa$^{1\,{\mbox{\LARGE$\star$}}}$\\
  $^1$Department of Astronomy, Faculty of Science, Kyoto University,
  Kyoto 606-8502, Japan\\
  $^2$Faculty of Education, Shiga University, 2-5-1 Hiratsu, Otsu,
  Shiga 520-0862, Japan}

\date{Accepted 1999 xxxx xx.
  Received 1999 February xx}

\pagerange{\pageref{firstpage}--\pageref{lastpage}}
\pubyear{1999}

\begin{document}

\maketitle

\label{firstpage}

\begin{abstract}
  A recent observation with the {\it Hipparcos} satellite and some
  numerical simulations imply that the interaction between an oblate
  halo and a disc is inappropriate for the persistence of galactic
  warps.  Then, we have compared the time evolution of galactic warps
  in a prolate halo with that in an oblate halo.  The haloes were
  approximated as fixed potentials, while the discs were represented
  by $N$-body particles.  We have found that the warping in the oblate
  halo continues to wind up, and finally disappears.  On the other
  hand, for the prolate halo model, the precession rate of the outer
  disc increases when the precession of the outer disc recedes from
  that of the inner disc, and vice versa.  Consequently, the warping
  in the prolate halo persisted to the end of the simulation by
  retaining the alignment of the line of nodes of the warped disc.
  Therefore, our results suggest that prolate haloes could sustain
  galactic warps.  The physical mechanism of the persistence of warp
  is discussed on the basis of the torque between a halo and a disc
  and that between the inner and outer regions of the disc.
\end{abstract}


\section{Introduction}
Many spiral galaxies, including our Galaxy, have warped discs which
resemble characteristic `cosmic integral signs'.  That is, the outer
disc lies above the inner disc plane on one side, and falls bellow
that plane on the other side.  Although the warping is often seen in
neutral hydrogen layers (Sancisi 1976; Bosma 1981), it is also
observed in stellar discs (van der Kruit \& Searle 1981; Innanen et
al.\ 1982; Sasaki 1987).  In the Milky Way, the stellar warp has been
detected not only for young stars (Miyamoto, Yoshizawa \& Suzuki 1988)
but for old stars (Porcel, Battaner \& Jim\'enez-Vicente 1997).  In
addition, the frequency of warped discs in spiral galaxies is
sufficiently large that at least half of spirals are warped both in H
{\sc i} discs (Bosma 1991) and in optical discs (S\'anchez-Saavedra,
Battaner \& Florido 1990; Reshetnikov \& Combes 1998).  These
observations imply that warps must persist for a long time unless they
are repeatedly excited.  It is true that some galaxies with warped
discs (e.g., M31, see Innanen et al.\ 1982) have nearby companions.
However, there do exist warped galaxies (e.g., NGC 4565, see Sancisi
1976) that have no nearby companions being supposed to be responsible
for the warp in the recent past.  In fact, Reshetnikov \& Combes
(1998) have revealed that about 21 out of 133 isolated galaxies are
warped like an integral sign.  This indicates that warps are not
necessarily caused by tidal interactions with other galaxies.

One explanation for isolated warped galaxies is the gravitational
torque of a halo acting on a disc that is `misaligned' to the
equatorial plane of the halo.  Such a tilted disc embedded in a halo
has intrinsic spin, so that it precesses like a top.  Since the
precessing rate is a function of radius, kinematical warps will wind
up and disperse in a short period of time.  Once the self-gravity of
the disc is taken into account, realistic warped configurations emerge
in which a disc precesses coherently like a solid body inside
axisymmetric haloes (Sparke 1984; Sparke \& Casertano 1988, hereafter
SC; Kuijken 1991).  Even if a tilted disc is formed in a halo as a
different shape from a warped mode, it will be finally turned into the
mode within a Hubble time (Hofner \& Sparke 1994).  However, according
to some numerical simulations (Dubinski \& Kuijken 1995; Binney, Jiang
\& Dutta 1998), the warping in oblate haloes is not retained
persistently, and so, disappears within a few dynamical times.  Thus,
the interaction between an oblate halo and a disc will be
inappropriate for long-lasting warps.

Recently, Smart et al.\ (1998) have extracted warp-induced motions in
the Milky Way by analysing the data obtained with the {\it Hipparcos}
satellite.  Then, they have found that the Galactic warp rotates in
the same direction as the Galaxy.  As shown by Nelson \& Tremaine
(1995), oblate haloes lead to the opposite sense of the warp
precession to the Galactic rotation, whereas prolate haloes make them
rotate in accordance with each other.  In addition, they have
demonstrated that in some cases, prolate haloes can excite warps.
Thus, prolate haloes are favourable to the explanation of Smart et
al.'s (1998) finding, if the motions that they found are attributed to
the interaction of the Galactic disc with an often assumed massive
halo.  Cosmological simulations, based on a cold dark matter scenario,
have also revealed that dark matter haloes surrounding individual
galaxies are highly triaxial and that the fraction of prolate haloes
is roughly equal to that of oblate haloes (Dubinski \& Carlberg 1991).
In spite of these circumstances mentioned above, prolate haloes have
somehow often been ignored in previous studies on warps.  Therefore,
we need to pay attention to the warp arising from prolate haloes.

In this paper, we examine how a warp is developed and evolves in
prolate haloes in comparison with that in oblate haloes.  As a first
step, we treat the haloes as external fixed potentials.  In Section 2,
we describe the models and the numerical method.  Results are
presented in Section 3.  In Section 4, we analyse our results and
explain them on the basis of a simplified model.  Conclusions are
given in Section 5.

\section{Models and Method}
We study the evolution of self-gravitating discs embedded in
axisymmetric haloes.  As shown by Nelson \& Tremaine (1995) and by
Dubinski \& Kuijken (1995), dynamical friction between a disc and a
halo plays an important role to precessing bending modes at least for
the inner region of the composite system.  However, the accurate
evaluation of dynamical friction would require a prohibitively huge
number of particles to represent the halo as well as the disc.
Otherwise, the disc will thicken owning to two-body relaxation
originating from Poisson fluctuations.  In fact, Dubinski \& Kuijken
(1995) reported the vertical disc thickening for a self-consistent
model with a particle disc, bulge, and halo.  As a result, warped
structures developed in the disc could not be distinguished from the
background particle distribution, which would lead us to an incorrect
conclusion about the longevity of the warp.  Then, we begin with rigid
halo models, as a first step, to unravel the effects of the halo shape
on the warp.

The density distribution of the halo is an axisymmetric modification
of Hernquist's models (Hernquist 1990) being suitable for spherical
galaxies and bulges.  Then, the halo density profile is represented,
in cylindrical coordinates, by
\begin{equation}
  \rho_{\rm h}\left(R,z\right)=\frac{M_{\rm h}}{2\pi a^2c}\frac{1}
{m\left(1+m\right)^3},
\end{equation}
where $M_{\rm h}$ is the halo mass, $a$ and $c$ are the radial and
vertical core radii, respectively, and
\begin{equation}
  m^2=\frac{R^2}{a^2}+\frac{z^2}{c^2}.
\end{equation}
The cumulative mass profile and potential of the halo are written,
respectively, by (Binney \& Tremaine 1987),
\begin{equation}
  M_{\rm h}\left(R,z\right)=M_{\rm h}\frac{m^2}{\left(1+m\right)^2},
\end{equation}
and
\begin{equation}
  \Phi_{\rm h}\left(R,z\right)=-\frac{GM_{\rm h}}{2}\int_0^\infty\frac{du}
{\left(a^2+u\right)\sqrt{c^2+u}\left[1+m\left(u\right)\right]^2},
\end{equation}
where
\begin{equation}
  m^2\left(u\right)=\frac{R^2}{a^2+u}+\frac{z^2}{c^2+u}.
\end{equation}

\begin{table}
  \caption{Parameters for $N$-body simulations.}
  \begin{tabular}{rcccccc}
    \hline          
    & $M_{\rm d}$ & $R_{\rm d}$ & $z_{\rm d}$ & $M_{\rm h}$ & $a$ & $c$\\
    \hline
    OBLATE  & 1.0 & 1.0 & 0.2 & 17.8 & 10.0 & 7.5\\
    PROLATE & 1.0 & 1.0 & 0.2 & 17.8 & 7.5  & 10.0\\
    \hline
  \end{tabular}
  \label{tab:para}
\end{table}

\begin{figure}
  \begin{center}
    \epsfxsize=0.8\hsize \epsfbox{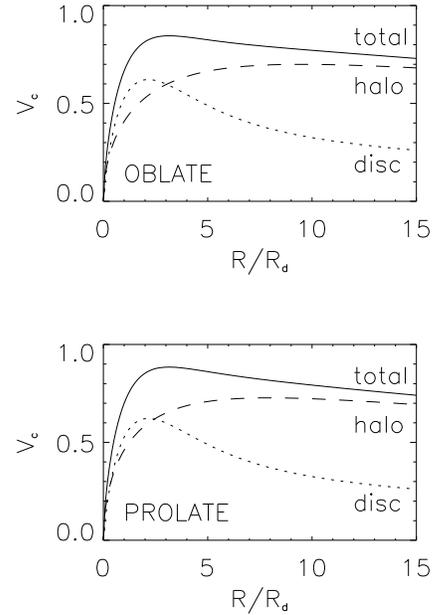}
  \end{center}
  \caption{Circular speed showing the contribution from the disc and halo to
    the total for the oblate (top) and prolate (bottom) halo models.}
  \label{fig:rc}
\end{figure}

As a realistic disc model, though a bulge component is not included,
we adopt an exponential density profile in the radial direction
(Freeman 1970) and an isothermal sheet approximation in the vertical
direction (Spitzer 1942) given by
\begin{equation}
  \rho_{\rm d}\left(R,z\right)=\frac{M_{\rm d}}{4\pi R^2_{\rm d}z_{\rm d}}
\exp\left(-\frac{R}{R_{\rm d}}\right){\rm sech}^2\left(\frac{z}{z_{\rm d}}
\right),
\end{equation}
where $M_{\rm d}$ is the disc mass, $R_{\rm d}$ is the disc
scale-length, and $z_{\rm d}$ is the disc scale-height.  The discs are
truncated radially at $15\;R_{\rm d}$ and vertically at $2\;z_{\rm
  d}$.  Following Hernquist's (1993) approach, we approximate the
velocity distribution of disc particles using moments of the
collisionless Boltzmann equation; the velocities are sampled from
Gaussian distributions with means and dispersions derived from the
Jeans equations.  The Toomre $Q$ parameter (Toomre 1964) used to
normalize the radial velocity dispersion is set to be 1.5 at the solar
radius, ${\rm R}_{\odot}=\left(8.5/3.5\right)\;R_{\rm d}$.  The $Q$
profile varies with radius and has a minimum at approximately
$R=2\;R_{\rm d}$, where the minimum $Q$ value is about 1.47.  To avoid
complications due to an extra component such as a bar, this rather
large $Q$ distribution is chosen so that the bar instability will not
occur in the disc.

We have ensured that the disc-halo system thus constructed is really
in equilibrium if the disc is initially placed in the equatorial plane
of the halo, because the density profile of the disc did not change
significantly over several orbital times.  However, the disc does not
necessarily form in the equatorial plane of the halo.  In fact, Katz
\& Gunn (1991) showed that the disc was misaligned to the symmetry
plane of the halo at an angle of typically 30 degrees at its birth.
Therefore, in our simulations the disc is initially tilted by 30
degrees with respect to the equatorial plane of the halo.

We employ a system of units such that the gravitational constant
$G=1$, the disc mass $M_{\rm d}=1$, and the exponential scale-length
$R_{\rm d}=1$.  The orbital period at the half-mass radius of the
exponential disc, $R\simeq 1.7\;R_{\rm d}$, is $13.4$ in our system of
units.  If these units are scaled to physical values appropriate for
the Milky Way, i.e., $R_{\rm d}=3.5\;{\rm kpc}$ and $M_{\rm d}=5.6
\times 10^{10}\;{\rm M_\odot}$, unit time and velocity are $1.31\times
10^7\;{\rm yr}$ and $262\;{\rm km\;s^{-1}}$, respectively.

The disc is represented by $100\;000$ particles of equal mass.  We
show the parameters of our models in Table \ref{tab:para}, and the
rotation curves of each model in Fig.~\ref{fig:rc}.  The halo mass is
determined so that the disc and halo masses within $3\;R_{\rm d}$ are
equal to each other.

The simulations are run with a hierarchical tree algorithm (Barnes \&
Hut 1986) using the GRAPE-4, a special-purpose computer for
gravitationally interacting particles (Sugimoto et al.\ 1990; Makino
et al.\ 1997).  We adopt an opening angle criterion, $\theta=0.75$.
Only monopole terms are included in the tree code.  The equations of
motion are integrated with a fixed time-step, $\Delta t=0.1$, using a
time-centred leapfrog method.  The Plummer softening length is
0.04$\;R_{\rm d}$, or in other words, 0.2$\;z_{\rm d}$.

\section{Results}
We stopped the simulations at $t=400$.  This time corresponds to about
30 orbital periods at the half-mass radius of the disc.  No bar
instability was found in the discs.  In either simulation, the total
energy was conserved to better than 0.2 per cent.

\begin{figure}
  \begin{center}
    \epsfxsize=\hsize \epsfbox{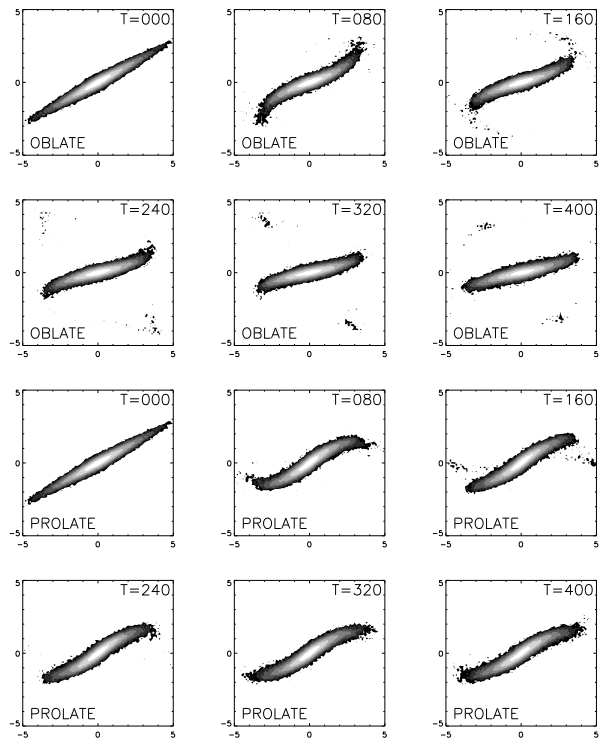}
  \end{center}
  \caption{Time evolution of the projected density distributions of the disc in
    the precessing frame.  The top (bottom) two rows correspond to the
    oblate (prolate) halo model.  The time is shown at the right-up
    corner of each frame.}
  \label{fig:edge}
\end{figure}

We measured the inclination and the longitude of ascending nodes of
the disc relative to the equatorial plane of the halo by calculating
the principal moments of inertia for the bound particles.  In
Fig.~\ref{fig:edge}, we then show the evolving density profiles from
an edge-on view of the discs in the precessing frame. In this frame,
an observer is always on that line of nodes of the disc which is
calculated for the particles within the half-mass radius of the disc,
$R\simeq1.7\;R_{\rm d}$.  The precession periods of the discs in the
oblate and prolate haloes evaluated with least-squares fits are
$T_{\rm ob}=266$ and $T_{\rm pr}=306$, respectively.  These values are
in excellent agreement with those predicted by linear theory [see
equation (21) of SC] which gives $T_{\rm ob}=266$ and $T_{\rm
  pr}=308$.

Hofner \& Sparke (1995) showed that warped configurations are
developed in oblate haloes.  We further find from Fig.~\ref{fig:edge}
that such configurations appear in the prolate halo as well as in the
oblate one.  In addition, Fig.~\ref{fig:edge} demonstrates that the
shape of warp depends on that of halo: using the terminology in SC,
the type I warp that bends upward away from the symmetric plane of the
halo is developed in the oblate halo, while the type II warp that
bends down toward the symmetric plane of the halo is developed in the
prolate one.  The warp in the oblate halo decayed and disappeared
almost completely by the end of the simulation, while the warp in the
prolate one persisted to the end.

\begin{figure}
  \begin{center}
    \epsfxsize=\hsize \epsfbox{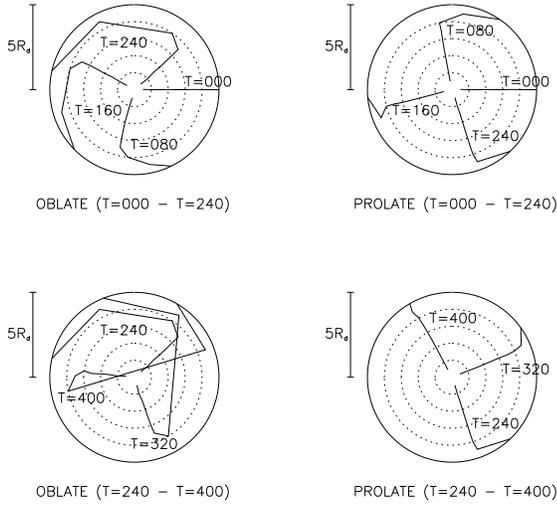}
  \end{center}
  \caption{Time evolution of the line of ascending nodes of discs on polar
    diagram. The left and right columns correspond to the oblate and
    prolate halo models, respectively. Concentric dotted lines are drawn
    at a radial interval of one sale length, with a solid line at five
    scale lengths.  The disc rotation is counterclockwise.}
  \label{fig:lon}
\end{figure}

To see the differential precession of the disc, we divided the
distance between the centre and $5\;R_{\rm d}$ evenly into 10 annuli,
and calculated the inclination and azimuthal angles of each annulus
which contains at least $2\;000$ particles.  Fig.~\ref{fig:lon} shows,
on the polar diagram, the line of ascending nodes of each annulus with
respect to the equatorial plane of the halo. There are two differences
between the oblate and prolate halo models.  One is the sense of the
precession; for the oblate halo the warp rotates in the opposite
direction to the disc rotation, while for the prolate one it rotates
in the same direction.  The other is the behaviour of the winding of
the warp; for the oblate halo the warp winds up tightly with time,
while for the prolate one the longitude of each annulus is kept almost
aligned.

We find from Fig.~\ref{fig:lon} that the inner region of the disc,
$R\la3\;R_{\rm d}$, precesses at almost a constant rate independent of
radius both in the prolate and oblate haloes.  Since the disc mass is
equal to the halo mass within $3\;R_{\rm d}$ for our models, the
self-gravity of the disc is dominant as compared to that of the halo
within such radius.  We thus understand that this behaviour arises
from the predominance of the self-gravity of the disc over that of the
halo, as shown by Lovelace (1998). On the other hand, the outer disc
$(R\ga3\;R_{\rm d})$ precesses at a different rate from radius to
radius.

\begin{figure}
  \begin{center}
    \epsfxsize=\hsize \epsfbox{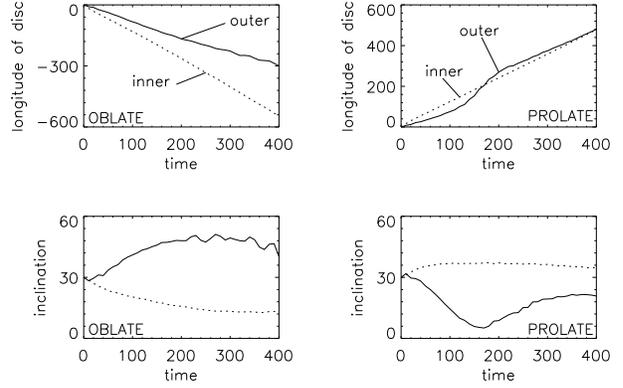}
  \end{center}
  \caption{Time evolution of the longitude of ascending node (top row) and the
    inclination (bottom row) of discs.  The solid and dashed lines show
    the outer and inner discs, respectively; the inner disc corresponds
    to a radial annulus between $1.5\;R_{\rm d}$ and $2\;R_{\rm d}$, and
    the outer disc to that between $4.5\;R_{\rm d}$ and $5\;R_{\rm d}$.}
  \label{fig:angle}
\end{figure}

To make clear the difference in precession at large radii between the
oblate and prolate halo models, we present in Fig.~\ref{fig:angle} the
time evolution of the longitude (top row) and that of the inclination
angle (bottom row) with respect to the equatorial plane of the halo
for the outer and inner discs which correspond to the annulus between
$4.5\;R_{\rm d}$ and $5.0\;R_{\rm d}$, and to that between
$1.5\;R_{\rm d}$ and $2.0\;R_{\rm d}$, respectively.
Fig.~\ref{fig:angle} shows that the precession rate of the outer disc
increases with decreasing inclination angle, and vice versa.  For the
prolate halo, at the beginning of the simulation, the inclination
angle decreased, and the precession rate increased.  In the subsequent
evolution, the longitude of the outer disc passed through that of the
inner disc as the inclination decreased.  After the passage of the
longitude of the outer disc, the inclination increased, and the
precession rate decreased.  For the oblate halo, on the other hand,
the inclination increased initially, and the precession rate
decreased.  The difference in longitude between the inner and outer
discs became larger with increasing inclination, and so, the warp
wound up with time as seen in Fig.~\ref{fig:lon}.

The difference in inclination angle between the inner and outer discs
in the oblate halo is larger than that in the prolate one at the end
of the simulation.  However, for the oblate halo, the warp disappeared
as seen in Fig.~\ref{fig:edge}, because the azimuthal angle of the
inner disc was different from that of the outer disc.

\section{Physical interpretation}
As found in the previous section, the prolate halo is plausible for
the maintenance of galactic warps in that the winding problem is
avoided.  However, there remains a question as to what makes the
difference in evolution of warps between the oblate and prolate halo
models.

Since Lovelace (1998) has shown that the self-gravity of the disc can
synchronize the precession rate in the inner region, we need to
explain the different behaviour of the warp in the outer region
between the oblate and prolate halo models.  Then, we simplify the
$N$-body models used in our simulations and construct a
three-component system consisting of an outer disc, an inner disc, and
an axisymmetric halo in order to pay special attention to the torque
between the inner and outer discs in addition to that from the halo.
Here, we approximate an outer disc as a ring.

\subsection{Simple model}
In this subsection, we solve the equation of motion for the outer ring
in order to examine whether the results found in the $N$-body
simulations can be reproduced.  For this purpose, we consider the
axisymmetric Binney (1981) potentials as models of the halo and the
inner disc,
\begin{eqnarray}
  \Phi_{\rm h}&=&\frac{1}{2}V^2_{\rm c,h}\ln\left(R^2_{\rm c,h}+R^2+\frac{z^2}
  {q^2_{\rm h}}\right),\\
  \Phi_{\rm d}&=&\frac{1}{2}V^2_{\rm c,d}\ln\left(R^2_{\rm c,d}+R^2+\frac{z^2}
  {q^2_{\rm d}}\right),
\end{eqnarray}
where $V_{\rm c}$ is the asymptotic circular velocity, $R_{\rm c}$ is
the core radius, and $q$ is the potential flattening with the
subscripts `h' and `d' denoting the halo and the disc, respectively.
The potentials of the halo and the inner disc are fixed.  The inner
disc is tilted by an angle of 30 degrees relative to the symmetric
plane of the halo and given a constant precession rate that is
calculated from linear theory (SC).  The dynamics of the outer ring
are solved on the basis of Euler's equation of motion for a rigid body
(Goldstein 1980).

The radius and angular speed of the outer ring, $\Omega$, are set to
be $5.0$ and $0.15$, respectively.  The parameters of this model are
summarized in Table \ref{tab:model}.  The values of the potential
flattening, $q_{\rm h}$ and $q_{\rm d}$, are adjusted to those which
are evaluated from the halo and disc models at $5\;R_{\rm d}$ employed
in the $N$-body simulations (see Table \ref{tab:para}).  We determine
the values of $V_{\rm h,d}$ and $V_{\rm c,d}$ in the same manner.

\begin{table}
  \caption{Parameters for three-component models.}
  \begin{tabular}{rcccccc}
    \hline
    & $V_{\rm c,d}$ & $R_{\rm c,d}$ & $q_{\rm d}$ & $V_{\rm c,h}$ &
    $R_{\rm c,h}$ & $q_{\rm h}$\\
    \hline
    OBLATE  & 0.5 & 1.0 & 0.8 & 0.7 & 1.0 & 0.9\\
    PROLATE & 0.5 & 1.0 & 0.8 & 0.7 & 1.0 & 1.1\\
    \hline
  \end{tabular}
  \label{tab:model}
\end{table}

\begin{figure}
  \begin{center}
    \epsfxsize=\hsize \epsfbox{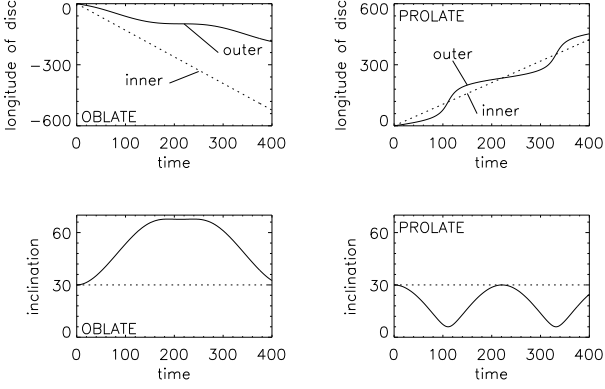}
  \end{center}
  \caption{Same as Fig.~\ref{fig:angle}, but for three-component models.}
  \label{fig:model}
\end{figure}

Fig.~\ref{fig:model} presents the time evolution of the longitude and
inclination of the outer ring, which corresponds to
Fig.~\ref{fig:angle} in the $N$-body simulations.  We can see that the
precession rate of the outer ring increases with decreasing
inclination angle with respect to the equatorial plane of the halo,
and vice versa.  Moreover, for the prolate halo the inclination angle
decreases at the beginning, while for the oblate one it increases.
Thus, for the prolate halo, the precession of the outer ring can pass
through that of the inner disc.  After the passage, the inclination
angle increases and the precession rate decreases.  This means that
the longitude of the inner disc and the outer ring remains almost
aligned.  Therefore, the behaviour of the outer ring is the same as
that seen in the $N$-body simulations.  Since we have found that the
simple model can reproduce the main properties of the $N$-body
simulations, we can rely on this model to figure out the physical
mechanism of warps in the oblate and prolate haloes as described
below.

\subsection{The precession}
\begin{figure}
  \begin{center}
    \epsfxsize=0.8\hsize \epsfbox{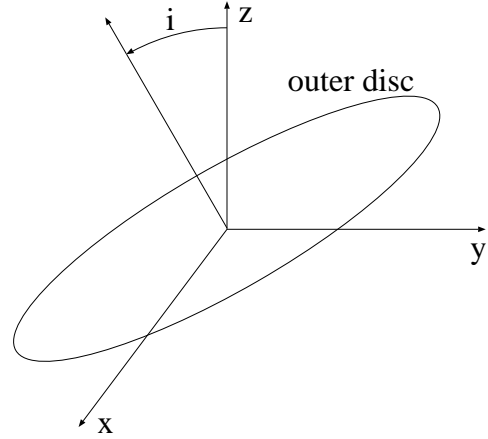}
  \end{center}
  \caption{Geometry for the outer disc. The line of node is along the $x$-axis,
    and the symmetry axis of the halo is along the $z$-axis.}
  \label{fig:geo}
\end{figure}

The precession of the outer ring is caused by the torque due to the
halo and the inner disc.  We take a coordinate system in which the
$z$-axis is along the symmetry axis of the halo, and the $x$-axis is
along the line of nodes of the outer ring (where the outer ring
intersects with the $z=0$ plane).  The $xy$-plane is in the inertial
frame.  The geometry is shown in Fig.~\ref{fig:geo}.

The $x$-component of the torque exerted by the halo on the outer ring
is given by
\begin{equation}
  T_{{\rm h},x}=-\int_0^{2\pi}rd\phi\lambda\left(F_{{\rm h},y}z-
    F_{{\rm h},z}y\right)\quad\lambda\equiv\frac{m}{2\pi r},
\end{equation}
where $\lambda$ is the line density of the outer ring, and $m$ is the
mass of the outer ring.  The $y$- and $z$-components, $F_{{\rm h},y}$
and $F_{{\rm h},z}$, of the force due to the halo per unit mass are,
respectively, written by
\begin{equation}
  F_{{\rm h},y}=-\Omega^2_{\rm h}y, \quad F_{{\rm h},z}=-\mu^2_{\rm h}z,
\end{equation}
where $\Omega_{\rm h}$ and $\mu_{\rm h}$ are the halo contributions of
the orbital and vertical frequencies, respectively.  The position
vector of a point on the outer ring is
\begin{equation}
  \mbox{\boldmath $R$}=\left(r\cos\phi,r\cos i\sin\phi,r\sin i\sin
\phi\right)\quad 0\le\phi\le 2\pi,
\end{equation}
where $i$ is the inclination angle of the outer ring with respect to
the equatorial plane of the halo.  Then, on the assumption that
$\Omega_{\rm h}$ and $\mu_{\rm h}$ are constant on the ring, we
obtain
\begin{equation}
  T_{{\rm h},x}=\frac{mr^2}{2}\left(\Omega^2_{\rm h}-\mu^2_{\rm h}\right)
  \sin i\cos i.
\end{equation}
The potential of an exponential disc in the outer region is
approximated as [see equation (2P-5) of Binney \& Tremain (1987)],
\begin{equation}
  \label{eq:disc}
  \Phi_{\rm d}\left(R,z\right)\simeq-\frac{GM_{\rm d}}{r}
  \left[1+\frac{3R^2_{\rm d}\left(R^2-2z^2\right)}{2r^4}\right].
\end{equation}
Thus, the $y$- and $z$-components of the force due to the inner disc
per unit mass are, respectively, given by
\begin{equation}
  F_{{\rm d},y}=-\frac{GM_{\rm d}}{r^2}\frac{y}{r}\left[1+\frac{15R^2_{\rm d}
      \left(R^2-2z^2\right)}{2r^4}-\frac{3R^2_{\rm d}}{r^2}\right],
\end{equation}
and
\begin{equation}
  F_{{\rm d},z}=-\frac{GM_{\rm d}}{r^2}\frac{z}{r}\left[1+\frac{15R^2_{\rm d}
      \left(R^2-2z^2\right)}{2r^4}+\frac{6R^2_{\rm d}}{r^2}\right].
\end{equation}
Provided that the line of nodes with respect to the equatorial plane
of the halo is aligned between the inner disc and the outer ring, the
$x$-component of the torque exerted by the inner disc is
\begin{equation}
  T_{{\rm d},x}=-\frac{mr^2}{2}\frac{9R^2_{\rm d}GM_{\rm d}}{r^5}
  \sin\delta\cos\delta,
\end{equation}
where $\delta$ is the inclination angle of the outer ring relative to
the inner disc plane, and its sign is positive when the inclination
angle of the inner disc is larger than that of the outer ring, and
vice versa.

Next, the absolute value of the total angular momentum of the outer
ring, $L$, is $mr^2\Omega$, where $\Omega$ is the orbital frequency,
if the precession rate is rather smaller than the orbital frequency.
The perpendicular component to the $z$-axis is $mr^2\Omega\sin i$.
The change in $L_x$ over an infinitesimally small time $\Delta t$ is
$mr^2\Omega\sin i\omega_{\rm p}\Delta t$, where $\omega_{\rm p}$ is
the precession rate of the outer ring.  Thus,
\begin{equation}
  \dot{L_x}=mr^2\Omega\sin i\;\omega_{\rm p}.
\end{equation}
This should be equal to the torque on the outer ring, so that the
precession rate is derived as
\begin{equation}
  \label{eq:pre}
  \omega_{\rm p}=\frac{\Omega^2_{\rm h}-\mu^2_{\rm h}}{2\Omega}\cos i-
  \frac{9R^2_{\rm d}GM_{\rm d}}{2 \Omega r^5}\frac{\sin\delta\cos\delta}
  {\sin i}.
\end{equation}

\begin{figure}
  \begin{center}
     \vspace*{7cm} \LARGE{fig7.gif}
  \end{center}
  \caption{A sketch of the type I (a) and the type II (b) warps seen edge-on.}
  \label{fig:pic}
\end{figure}

If the warp is of the type I shape such as Fig.~\ref{fig:pic}a, $i$ is
nearly equal to $\delta$.  It follows from equation (\ref{eq:pre})
that the precession rate is proportional to $\cos i$.  In this case,
the second term of equation (\ref{eq:pre}) is negative because
$\delta$ is positive, and the first term is also negative because
$\Omega_{\rm h}<\mu_{\rm h}$ for the oblate halo.  Therefore, the
precession rate increases with increasing inclination as seen in
Fig.~\ref{fig:angle}, and the difference in longitude between the
outer ring and the inner disc becomes larger with time.  If the warped
disc is of the type II shape such as Fig.~\ref{fig:pic}b, $\delta$ is
negative.  Taking into consideration $\Omega_{\rm h}>\mu_{\rm h}$ for
the prolate halo, both terms in equation (\ref{eq:pre}) are positive.
Moreover, $i$ will become smaller than $|\delta|$ with decreasing
inclination, so that the second term of equation (\ref{eq:pre}) is
dominant.  Therefore, the precession rate increases with decreasing
inclination, and vice versa as seen in Fig.~\ref{fig:angle}.

\subsection{The inclination}
The time evolution of the inclination is also explained simply by the
torque on the outer ring.  The geometry is the same as that used in
the previous subsection except that the $xy$-plane is in the
precessing frame of the outer ring.

The $y$-component of the torque, $T_y$, changes that of the angular
momentum, $L_y=-mr^2\Omega\sin i$, and so, affects the inclination
angle, $i$.  At the beginning, $L_y$ suffers no change from the torque
due to the halo because of the axisymmetric nature of the halo.  If
the longitude of the outer ring is the same as that of the inner disc,
we obtain $T_y=0$, and so, the inclination remains unchanged.  If the
longitude of the outer ring is smaller than that of the inner disc,
i.e., a warp like a trailing spiral, $T_y$ becomes positive.  Hence,
$L_y$ increases, so that the inclination angle of the outer ring
decreases.  Since the torque on the inner disc is the opposite
direction to that of the outer ring according to Newton's third law,
the inclination of the inner disc increases.  As a result, the type II
warp is generated.  Similarly, a warp like a leading spiral leads to
the type I warp.

If the precession rate in the prolate halo decreases with radius, the
warped configuration becomes similar to a trailing spiral because the
sense of the precession is the same direction as the disc rotation.
Consequently, the type II warp is produced, which leads to the
situation where the inclination of the outer ring decreases, and the
precession rate increases.  In the subsequent evolution, the longitude
of the outer ring passes through that of the inner disc.  At the
point, the warped configuration turns into that similar to a leading
spiral, which, this time, leads to the situation where the inclination
increases and the precession rate decreases.  This kind of
self-regulation enables the warped disc to avoid the differential
precession.  For the oblate halo, however, once the precession of the
outer ring recedes from that of the inner disc, the difference in
precession rate continues to increase, so that the warp winds up
tightly.

\subsection{Comparison with discrete bending modes}
\begin{figure}
  \begin{center}
    \epsfxsize=0.8\hsize \epsfbox{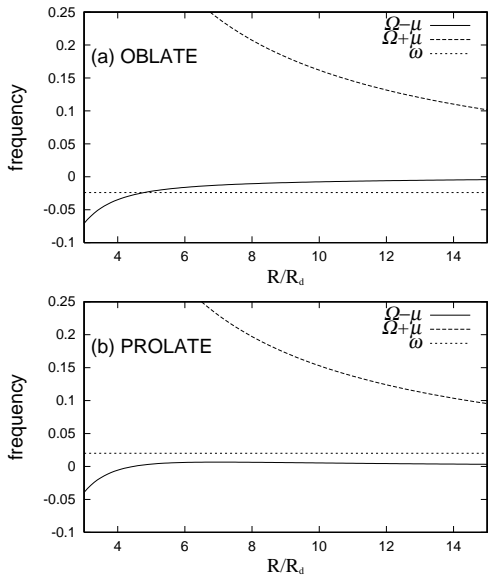}
  \end{center}
  \caption{Plots of the frequencies, $\Omega-\mu$, $\Omega+\mu$, and $\omega$
    for the oblate halo model (a) and the prolate halo model (b).  In
    the calculation of $\Omega-\mu$ and $\Omega+\mu$, the approximated
    form of equation (\ref{eq:disc}) is used for the disc potential.}
  \label{fig:omega}
\end{figure}

Our oblate halo model cannot sustain the warped disc, though SC found
that long-lived warps do exist in some halo models.  This discrepancy
is considered to originate from the radial extent of a disc.  SC
showed that a discrete mode with an eigenfrequency $\omega$ may exist
only if $\Omega-\mu <\omega <\Omega+\mu$ is satisfied at the edge of
the disc, where $\Omega$ and $\mu$ are the orbital and vertical
frequencies, respectively.  For oblate haloes, $\Omega-\mu$ and
$\omega$ are negative.  Since $\Omega-\mu$ tends to zero with radius,
there is some radius where $\omega$ is equal to $\Omega-\mu$.  Beyond
such a radius, the condition of the existence of discrete modes is
violated.  According to SC, bending modes become continuous in
frequency $\omega$, if a warped disc embedded in an oblate halo
extends beyond the radius at which $\omega=\Omega-\mu$ holds.
Consequently, such continuous modes will be propagated with a group
velocity and disappear (Hunter \& Toomre 1969).  As is found from
Fig.~\ref{fig:omega}a, such a resonance radius emerges at
$R\simeq4.8\;R_{\rm d}$ in our oblate halo model.  This implies that
there would be no discrete bending mode and that a warped
configuration would be coerced to disperse.  Thus, the disappearance
of the warping for the oblate halo could be due to the existence of
the resonance.  On the other hand, in our prolate halo model, the
condition of $\Omega-\mu < \omega < \Omega+\mu$ is exactly satisfied
within the truncated radius, 15 $R_{\rm d}$, as shown in
Fig.~\ref{fig:omega}b.  Therefore, a discrete mode can exist in our
prolate halo model.

Our adopted disc model is nothing special in the sense that the
observed light distributions of galactic discs are well-described by
an exponential law, though a constant mass-to-luminosity ratio
throughout the disc is assumed.  Our simulations suggest that as long
as a disc is not truncated abruptly, real galactic discs would not
have a discrete bending mode if surrounding haloes are oblate.

\section{Conclusions}
In this paper, we have examined the time evolution of warped discs in
the oblate and prolate haloes using $N$-body simulations.  The haloes
were represented by fixed external potentials in which
self-gravitating discs were embedded.  Then, we have found the warped
configurations both in the oblate and prolate haloes.  While the
warping in the oblate halo continued to wind up with time and finally
disappeared, the warping in the prolate halo survived to the end of
the simulation by regulating the line of nodes of the warped disc to
be straight.  We have shown that this difference in winding between
the oblate and prolate haloes can be attributed to the gravitational
torque between the inner and outer discs.

Observationally, some galaxies show the straight line of nodes of the
warp within the Holmberg radius, beyond which the warp is traced as a
leading spiral (Briggs 1990).  Others show that the line of nodes is
delineated as a trailing spiral (Christodoulou, Tohline,
Steiman-Cameron 1988; Bosma 1991).  These observations appear to
favour the view that we just witness the different phases of the
evolving warped discs in prolate haloes, because only warped discs in
prolate haloes can change the spirality of the line of nodes according
to our simulations.

Putting together our simulations and the observations mentioned above,
we can infer that warps are formed and maintained in prolate haloes.
However, our fixed halo models are quite simplified.  In particular,
such models cannot include the effect of dynamical friction between
the warped disc and the halo.  Dubinski \& Kuijken (1995) and Nelson
\& Tremaine (1995) have shown that dynamical friction plays an
important role to precessing discs.  Therefore, we will need the
simulations of warped discs embedded in live haloes to determine the
precise evolution of warps, though a huge number of particles will be
required to incorporate the effect of dynamical friction accurately
into such simulations, and to avoid disc heating due to an
insufficient number of halo particles.  This line of investigation is
in progress (Ideta et al.\ in preparation).

\section*{Acknowledgments}
We are grateful to Prof.\ S.\ Inagaki for his critical reading of the
manuscript.  We thank E.\ Ardi and Y.\ Kanamori for useful
discussions, and Dr.\ J.\ Makino and the anonymous referee for
valuable comments on our paper.  MI thanks Dr.\ J.\ Makino for giving
him an opportunity to use the GRAPE-4 and providing him with a tree
code available on it.  MI is also indebted to A.\ Kawai for his
technical advice on the use of the GRAPE-4.  TT and MT acknowledge the
financial support from the Japan Society for the Promotion of Science.

\bsp

\label{lastpage}

\end{document}